\documentclass[aps,prd]{revtex4}
\usepackage{mathrsfs}
\usepackage{tikz}
\usepackage{amsmath}
\usepackage{txfonts}

\usepackage{graphicx}
\graphicspath{{./images/}}

\newcommand{\be}{\begin{equation}}
\newcommand{\ee}{\end{equation}}
\newcommand{\ba}{\begin{eqnarray}}
\newcommand{\ea}{\end{eqnarray}}

\newcommand{\bo}{\raise-1mm\hbox{\Large$\Box$}}
\begin{document}

\title{Quantum entropy, countable sets, metric and connection}

\author{Kaushik Ghosh\footnote{E-mail kaushikhit10@gmail.com}}
\affiliation{Vivekananda College,University of Calcutta, 
269, Diamond Harbour Road, Kolkata - 700063, India}

\maketitle

\section*{Abstract}

We compare two methods for evaluating the cardinality of the Cartesian product $N \times N$ of the set of natural numbers $N$. The first is used to explain the thermodynamics of black body radiation by using convergent functions on 
$N \times N$. The cardinality of $N \times N$ enters through the partition function, internal energy and entropy for every macrostate given by a normal mode of electromagnetic wave. Here, $N \times N$ is assigned a greater cardinality than $N$. The second method was devised in analysis to count the rational numbers by using divergent functions on $N \times N$. Here, $N \times N$ is not assigned a greater cardinality than $N$. In this article, we show that the experimentally confirmed first approach is mathematically more consistent with the definition of the real line and foundations of topology. It also provides a quantitative measure of the cardinality of $N \times N$ relative to that of N. Similar arguments show that the set of rational numbers is not countable. This article suggests that the axiom of choice may become a useful technique to prove the existence theorems for connection and metric on the spacetime manifold than the usual application of second-countability.

\vspace{0.8cm}

{\bf MSC:} 58A05, 81P17, 03E10, 53B05, 46C05 



\vspace{0.8cm}

Key-words: quantum entropy, countable sets, stereographic projection, metric, connection, axiom of choice



\section*{1. Introduction}

In this manuscript, we will discuss the cardinality of $N \times N$ and the rational numbers from geometrical and physical perspectives [1,2]. We will demonstrate that convergent functionals like internal energy and quantum entropy are better suited to evaluate the cardinality of $N \times N$, the Cartesian product of the set of natural numbers $N = \{ 0,1,2,...\},$ [3,4]. Here, we mention some notations and definitions that will be used in this article. $R$ represents the set of real numbers. The set of positive integers $\{ 1,2,3...\}$ is denoted by ${Z}_{+}$, [5]. The set of all integers including zero is given by $Z$. ${Q}_{+}$ represents the set of positive rationals. We will say two sets are \textit{isomorphic} when there is a \textit{bijection} between the two. The equivalence class of sets isomorphic to a given set $A$ will be represented as  $[A]$. The cardinality of a set $A$ will be given by $|A|$ and the absolute value of a variable $x$ will be given by $||x||$.

The cardinalities of Cartesian products like $N^n$ and $Z_+^n$, $n \in Z_+$, enter in mathematical physics in two ways. The first is quantum statistical mechanics where the set of microstates for many thermodynamic systems are isomorphic to $N^n$ or $Z_+^n$. A well known example is the black body radiation where the set of microstates 
associated with every macrostate is given by $N \times N$. The macrostates are normal modes of electromagnetic wave in an isothermal enclosure each having two possible transverse polarizations. Here, a well-defined probability distribution on $N \times N$ is used to construct different thermodynamic variables like internal energy and entropy. Experiments imply that $|N \times N| = |N|^2 > |N|$. This also leads us to conclude that these are extensive thermodynamic variables [3]. In the next section, we will elaborate the corresponding counting procedure. We will briefly discuss the significance of cardinal numbers [5,6] for microcanonical ensembles.
$|N \times N|$ is also required to find the cardinality of rational numbers [5,6]. The rational numbers can be used to construct a covering of $R^n$ that is frequently used in analysis, topology and differential geometry [5,7]. In this case it is claimed that $|Z_+ \times Z_+| < |Z_+|$ and the method used to support this can be extended to conclude that $|N \times N| < |N|$, [5,6]. This method uses injective functions on $Z_+ \times Z_+$ and $N \times N$ that become divergent in the limit when any one of the argument approaches infinity [5,6]. $|Z_+ \times Z_+| < |Z_+|$ is then used to show that the rational numbers are countable [5,6]. Countability of the rational numbers is a special case of the continuum hypothesis which states that there exists no set having greater cardinality than $Z_+$ and lesser cardinality than $R$ [5,6]. Thus, we have a contradiction between the two approaches to evaluate $|N \times N|$ that needs to be resolved while we apply topology and differential geometry to develop formal theories of classical and quantum gauge fields, including gravity. The latter includes introduction of metric and connection in the spacetime manifold as an important example [7,8]. Countability of the rational numbers leads to the concept of completely separable (second-countable) topological spaces where one can introduce metric without using the axiom of choice [7,8]. We will discuss this further in the last section.

In this article, we will demonstrate that the experimentally confirmed method adapted in quantum statistical mechanics  to evaluate $|N \times N|$ is mathematically more consistent with the definition of real line and foundation of differential topology compared to the method used to show that $|N \times N| < |N|$. It also provides an actual act of counting and a quantitative  measure of $|N \times N|$ relative to $|N|$. The latter is not possible in the method that claims $|N \times N| < |N|$. We will discuss the cardinality of $N \times N$ in the second section.
Eqs.(6,7,8,10,13) in the second section demonstrate an experimentally observed counter example to the claim that $|N \times N| < |N|$. The cardinality of the rational numbers will be discussed in the last section. In many cases, thermodynamic entropy gives us the scope to use ordinary real numbers to compare cardinalities given by cardinal numbers [5,6]. This article suggests that the \textit{axiom of choice} could be 
the more rigorous technique to prove the existence of connection and metric in a topological manifold which otherwise use second-countability.

\section*{2. Black body radiation and cardinality of $N \times N$}

The set of real numbers $R$ forms an algebraic field with an order relation $<$ satisfying the following
order properties [7]:

\vspace{0.5cm}

\noindent{${O_{1}}:$ the order relation $ < $ has the least upper bound property. Every nonempty subset
	$A$ of $R$ that is bounded above has a least upper bound,}

\vspace{0.5cm}

\noindent{${O_{2}}:$ if x $<$ y, then there exists an element z such that x $<$ z and z $<$ y, where $x,y,z \in R$.}

\vspace{0.5cm}

\noindent{The real numbers with the above order properties form a linear continuum 
	and coordinatize the real line by the completeness axiom [5,9,10]. We can introduce
	the standard topology of open intervals $(a,b)$ in $R$ where $a,b \in R$ and $a < b$, [5,7]. This topology coincides with the order topology derived from the order relation of $R$ and satisfy the Hausdorff condition
	due to the order property $O_{2}$ [7]. We express $R$ as $(- \infty, \infty)$ with the understanding
	that the limit $x \rightarrow \infty, ~ x \in R$ means $x$ can be increased indefinitely without any upper bound [5]. This implies that $z$ given by: $z = x + \epsilon, ~ \epsilon > 0,$ ceases to exist as a real number in the limit $x \rightarrow \infty$. Otherwise, we can choose the g.l.b of the set $\{z = x + \epsilon ~|~ \epsilon > 0, x \rightarrow \infty \}$ as the l.u.b of $R$ to have a connected set. This is similar but not same as the Dedekind cut axiom [7]. We can denote this hypothetical real number by $\infty$. However, this hypothetical real number together with the hypothetical set of real numbers $\{z = x + \epsilon ~|~ \epsilon > 0, x \rightarrow \infty \}$ violate the original assumption that the limit $x \rightarrow \infty, ~ x \in R$ means that $x$ can be increased indefinitely without any upper bound. We summarize the above discussions as the fundamental definition of the limit $x \rightarrow \infty$:}

\vspace{0.3cm}

\textbf{Definition 1-1.} The real number: $z = x + \epsilon, ~ \epsilon > 0,$ ceases to exist in the limit $x \rightarrow \infty$.

\vspace{0.3cm}

\noindent{In other words, the closure property does not hold for $z = x + \epsilon, ~ \epsilon > 0$ in the limit $x \rightarrow \infty$. A similar discussion will remain valid for $w = x - \epsilon, ~ \epsilon > 0$ in the limit $x \rightarrow - \infty$. We have $b = a + \epsilon ~(> 0)$, for any pair of finite $a$ and $b$ with $a < b$. Thus, we may also use the order relation $<$ in $R$ and corresponding order properties $O_1$ and $O_2$ to state that the set $\{z ~| ~z > x\}$ ceases to exist as a set of real numbers in the limit $x \rightarrow \infty$. Physically, taking the limit $n \rightarrow \infty$ for the interval $(-n,n)$ means we have included all real numbers and no real number is available to represent the elements of $\{z ~| ~z > x, x \rightarrow \infty \}$. Such a set cannot be used to assign values to any physical variable. As an analogy, the function $f: (-1,1) \rightarrow (-1,1)~| f(x) = x + \epsilon, ~ 0 < \epsilon < 1$, does not exist because $x + \epsilon \geq 1$ for $x \geq 1 - \epsilon$. Before we proceed to stereographic projection in the following, we note that the definition: $R = (- \infty, \infty)$ also means that $x + \epsilon \in R$ for any pair of finite values of $x, \epsilon \in R$. This is useful to discuss various geometric transformations in $R^m$ where $m$ is a positive integer and does not remain valid for any $U = (-n,n),$ where $n \in Z_+$ is finite or bounded and $x, \epsilon \in U$.}

Definition 1-1 lies at the heart of topology and differential geometry. We illustrate its application in the stereographic projection from $S^1$ to $R$ given below:

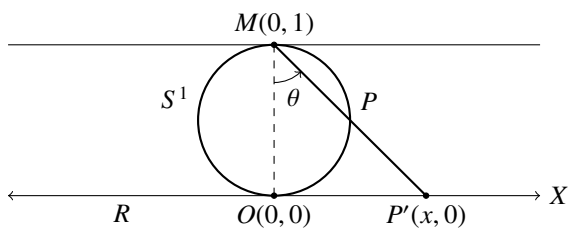
\vspace{0.2cm}

\begin{tikzpicture}
	
	\draw[<->] (-3.5,0) -- (3.5,0) node[right] {$X$};
	\draw[-] (-3.5,2) -- (3.5,2) node[right] {};
	
	\draw[thick] (0,1) circle (1cm);
	
	\filldraw (0,0) circle (1pt) node[below] {$O(0,0)$};
	\filldraw (0,2) circle (1pt) node[above] {$M(0,1)$};
	\filldraw (2,0) circle (1pt) node[below] {$P'(x,0)$};
	\draw (-2,0) node[below] {$R$};
	
	\draw[thick] (2,0) -- (0,2) node[midway, above right] {$P$};
	
	\draw[dashed] (0,0) -- (0,2);
	
	\draw[->] (0,1.5) arc[start angle=-90,end angle=-45,radius=0.5cm];
	\node at (0.25,1.3) {$\theta$};
	
	\node at (-1.3,1.3) {$S^1$};
	
\end{tikzpicture}

\vspace{0.2cm}

\noindent{Here, $\theta \in (-\pi/2 , \pi/2)$ and $P$ can be mapped to $P'$ with $x = \tan(\theta)$. We can use $x$ as the stereographic coordinate of $P$ under the given stereographic projection. We can't assign a stereographic coordinate to $M$ because two parallel straight lines in the Euclidean plane never intersect and there is no real number: $z = x + \epsilon, ~ \epsilon > 0$ in the limit $x \rightarrow \infty$ by Definition 1-1. Geometrically, there exists no positive real number that is not present on the positive $X$-axis. This leads us to conclude that $S^1$ contains more points than $R$ and gives a one-point compactification of $R$ [7]. $M$ can be denoted by $\infty \notin R$ [5]. We need at least two copies of $R$ to coordinatize $S^1$ when we use stereographic projections. We note that the stereographic projection discussed here gives a homeomorphism between the open interval $I = (-\pi/2, \pi/2)$ and $R$. We need a point $M \notin I$ to construct a circle from $I$. Thus, the stereographic projection demonstrates geometrically that homeomorphic sets have identical cardinality (Theorem 1-1). In this context, we note that the one-to-one transformation from $S^1$ to $(-\pi/2, \pi/2] \subset R$ used to coordinatize $S^1$ is not a continuous function [7] when the topologies on $S^1$ and $(-\pi/2, \pi/2]$ are induced from the topology given by open disks in $R^2$. The inverse image of any open set of $(-\pi/2, \pi/2]$ containing $\pi/2$ is not an open set of $S^1$. Thus, $S^1$ and $(-\pi/2, \pi/2]$ have different topological properties [7]. As an example, $S^1 \subset R^2$ is compact but $(-\pi/2, \pi/2] \subset R^2$ is not compact (Theorem 6.3, [5]). $S^1 - M$ and $R$ are homeomorphic. They are open and noncompact in the induced topology from $R^2$. In passing, we note that the limit $z = x - \epsilon, x \rightarrow \infty$ with $\epsilon > 0$ corresponds to $P[\theta - \delta\theta], \theta \rightarrow \pi/2$ on the $S^1$. $\delta\theta$ can be obtained using $x = \tan(\theta)$.}

We demonstrate that $|Q_+|$ can be greater than $|N|$ before we delve in the formal discussion on $|Q_+|$ and $|N|$. We consider the collection of open sets $\mathcal{C} = \{ (n - 2, n) | n \in Z_+ \}$. $\mathcal{C}$ covers the interval $(-1, \infty)$, [5]. There is only one integer, $(n - 1)$, in each $\mathcal{C}_n = (n - 2, n), ~ n \in Z_+ $. We consider the collection of set of rationals given by: $\mathcal{A} =  \{ n - {3 \over 4}, n - {1 \over 2} | ~n \in Z_+ \},$ such that $\mathcal{A}_n =  \{ n - {3 \over 4}, n - {1 \over 2} \} \subset \mathcal{C}_n$. We find that each $\mathcal{A}_{n}$ contains two rationals that are not integers and the corresponding $\mathcal{C}_{n}$ contains one integer. We also have $\mathcal{A}_k \cap \mathcal{A}_l = \emptyset, ~ \forall ~ k,l \in Z_+, k \neq l$ and $\bigcup_{n \in Z_+} \mathcal{A}_n \subset Q_+$. Thus, we find $|Q_+| > |N|$. A geometrical illustration of $(n - \epsilon), ~0 < \epsilon < 1$, in the limit $n \rightarrow \infty$ is given above.

We now state two theorems used to discuss the cardinality of a set. The first is the $Schr\ddot{o}der-Bernstein$ theorem 
which states [5,6]:

\vspace{0.3cm}

\textbf{Theorem 1-1} ($Schr\ddot{o}der-Bernstein$). If there are injections $f: A \rightarrow B$ and $g: B \rightarrow A$, then $A$ and $B$ have identical cardinalities.

\vspace{0.3cm}

\noindent{We note that isomorphic sets have the same cardinality. The $Schr\ddot{o}der-Bernstein$ theorem  remains valid even when $A$ and $B$ are not countable. The second theorem gives the condition for countability of a set and states [5]:}

\vspace{0.3cm}

\textbf{Theorem 1-2}. \textit{Let $B$ be a nonempty set. Then the following are equivalent:}

\vspace{0.1cm}

\noindent{(1) \textit{There is a surjective function $f: {Z_{+}} \rightarrow B$.}}

\noindent{(2) \textit{There is an injective function $g: B \rightarrow {Z_{+}}$.}}

\noindent{(3) \textit{$B$ is countable.}}

\vspace{0.2cm}

\noindent{The proof can be found in [5]. Here, $Z_{+}$ is assumed to be the largest countable set [5]. Thus, by countable we mean $B$ is either isomorphic to a proper subset of $Z_+$ or $B$ is isomorphic to $Z_{+}$ itself. In the later case, we have a bijection between $B$ and $Z_{+}$. They have the same cardinality by the $Schr\ddot{o}der-Bernstein$ theorem.}

	We conclude this section with a few discussions on quantum entropy and the cardinality of $N \times N$. In statistical mechanics, cardinality of a set enters through the partition function that serves as the normaliztion constant of a probability distribution over a set of relevant  microstates.  Internal energy and von Neumann quantum entropy, referred to simply as entropy henceforth, give a measure of the size of the underlying set of microstates associated with a given statistical system. We will discuss the partition function, internal energy and entropy of black body radiation. In this case, each of the three variables provides an experimentally confirmed method to compare the cardinality of $N \times N$ with that of $N$ itself. Here, we consider photons with two polarizations in a normal mode of electromagnetic wave. The normal modes are given by the members of a complex basis with periodic boundary conditions [3,11]. Corresponding microstates are Fock states [3,11] given by: $\Omega = \{ |n_1 , n_2> | ~n_1 , n_2 \in N \}$. $\Omega$ can be represented by $N \times N$ and we find: $|\Omega| = |N \times N|$. Partition function, entropy and internal energy are defined using convergent functionals on probability distributions over $\Omega$ and give a measure of: $|\Omega| = |N \times N|$.

	A photon confined in an isothermal cube of volume $L^3$ with frequency $\nu$ has the following energy and momentum [3]:
	
	\be
	E = h \nu = |\vec{p}|c, ~ \vec{p} = {{h \vec{q}} \over L} = \hbar \vec{k}, ~ \vec{q} = (q_1 , q_2, q_3), 
	~ q_i \in Z
	\ee
	
	\noindent{where $h$ is the Planck's constant with value $6.63 \times 10^{-34} J.s$ and $\hbar = h/2\pi$ [3]. Such a set of photons give an electric field: $\vec{E}(\vec{r}, t) = \vec{\mathcal{E}} \exp[{{2\pi i} \over h}({\vec{p}.\vec{r}} - Et)] = 
	\vec{\mathcal E} \exp ({\vec{k}.\vec{r}} - \omega t)$ with periodic boundary conditions on $\vec{E}(\vec{r}, t)$ and $\vec{\mathcal E}. \vec{k} = 0$. This gives two different polarizations $\vec{\mathcal{E}_1}$ and $\vec{\mathcal{E}_2}$ of the electric field that can be related to two different spin states of the photons [3,11]. 
	The number of photons associated with an electromagnetic field with wave vector $\vec{k}$ and polarization ${\vec{\mathcal E}_i}$ can take any value from $N$. Thus the microstates are given by: $\Omega = \{ |n_1 , n_2> | ~n_1 , n_2 \in N \}$, where the macrostates are given by the allowed values of $\vec{k}$ irrespective of polarization. The energy of the electromagnetic field in the state $|n_1 , n_2>$ with wave vector $\vec{k}$ is given by:}

\be
\xi \{ \vec{k}, n_i \} = h \nu (n_1 + n_2), ~ n_i \in N
\ee

\noindent{We now construct the partition function for the photon gas as a canonical ensemble (a grand canonical ensemble with vanishing chemical potential [3]) in thermal equilibrium with the enclosure at a temperature $T$. We first define a convergent function on $N \times N$ given by:}

\be
g_{\vec{k}}(n_1,n_2, \beta) = {\prod_{i = 1,2}}\exp(-{\alpha}{n_{i}})
= {\prod_{i = 1,2}}{[{\exp(-{\alpha})}]^{n_{i}}}, ~ \alpha = {{h \nu} \over {{k_B}T}} = \beta h \nu > 0
\ee

\noindent{where $k_B$ is the Boltzmann constant whose value is $1.38 \times 10^{23} J/K$ [3]. This is related to the probability of finding the system in the state $|n_1,n_2>$. We assume that $p(|n_1 , n_2>; \beta) = p(|n_1 >; \beta) p(|n_2>; \beta), ~n_i \in N$, where $p(|n>; \beta)$ is the probability of finding $n$ photons with single polarization given by either $\vec{\epsilon}_1$ or $\vec{\epsilon}_2$ at a temperature $T$. We note that this probability is independent of the polarization. This expression indicates that finding $n_1$ photons with polarization $\vec{\epsilon}_1$ and $n_2$ photons with polarization $\vec{\epsilon}_2$ are statistically independent events [3]. For $\nu = 0$, we consider only one state $|0,0>$ with probability one. We next consider the functional $\mathcal{Z}$ on $N \times N$ that give the relevant partition function for the photon gas with wave vector $\vec{k}$ at a temperature $T$:}

\ba
\mathcal{Z}(\vec{k}, \beta) & = & {\sum_{n_1}} {\sum_{n_2}} {g(n_{1}, n_{2}, \beta)} , ~~~ {n_{i}} \in N \\ \nonumber
~ & = & {\sum_{n_1}}\exp(-{\alpha}{n_{1}}) \Big[{\sum_{n_2}} \exp(-{\alpha}{n_{2}}) \Big ] \\ \nonumber
& \rightarrow & {({{e^{\alpha}} \over {e^{\alpha} - 1}})({{e^{\alpha}} \over {e^{\alpha} - 1}})} = {({{e^{\beta h \nu}} \over {e^{\beta h \nu} - 1}})^2} 
\ea	
	
\noindent{where we have first considered the partial sum of first $n + 1$ terms of the series
	${\sum}_{k}{\exp{(-\alpha k})}, ~ k \in N$. We express it as:}

\be
\sigma (n) = {1 \over {1 - r}} - {{r^n \over {1 - r}}} + {r^n}, ~~~~ r = \exp{(- \alpha)}.
\ee

\noindent{The sum over $N$ is obtained by taking the limit $n \rightarrow \infty$.
	The functions $r^n:(0,1) \rightarrow R$ converge to $0$ when $n \rightarrow \infty$,
	and $\mathcal{Z}$ converges to ${({{e^{\alpha}} \over {e^{\alpha} - 1}})^2}$. The probability for finding the system in state $|n_1,n_2>$ is given by:}

    \be
    p(|n_1,n_2>; \beta) = {g_{\vec{k}}(n_1,n_2, \beta) \over \mathcal{Z}(\vec{k}, \beta)}.
    \ee
    
    \noindent{Thus, the partition function $\mathcal{Z}(\vec{k}, \beta)$ gives the normalization constant 
    	of the distribution given by $g_{\vec{k}}(n_1,n_2, \beta)$. Thermodynamic averages can be obtained from the Fock states by using the following expression [11]:
    	
    	\[ {({A}(\vec{k}))}_{\beta} = \sum_{n_1 , n_2}  p(|n_1,n_2>; \beta) <n_1 , n_2|{\hat A}|n_1 , n_2>, ~ n_i \in N, \] 
    	where ${\hat A}$ is an operator. This expression can also be considered as the 
    	statement of completeness of the Fock space $\Omega$. A general macroscopic $(A)_{\beta}$ is obtained by superposing ${({A}(\vec{k}))}_{\beta}$ with different $\vec{k}$'s. $\Omega$ is also required to construct the coherent states [11]. Coherent states are required to construct states with non-vanishing expectation values for the electromagnetic fields themselves [11].}

    	We define the partition function for photons with only one of the two polarizations, \textit{i.e}, for either $\{|n_1> |~ n_1 \in N \}$ or $\{|n_2>|~n_2 \in N \}$ by: $z(\vec{k}, \beta) = ({{e^{\alpha}} \over {e^{\alpha} - 1}}) = \sqrt{\mathcal{Z}(\vec{k}, \beta)}$.  $z(\vec{k}, \beta)$ is a sum over $N$ with positive weights and gives a measure of $|N|$. Before we proceed further, we note that there can exist convergent probability distributions on $N$ for which corresponding partition function $z \leq 1$. An example is the partition function for phonons which are quanta of lattice vibrations in crystalline solids [3]. Here, $n_i$ in Eq.(3) is replaced by $n_i + \frac{1}{2}, ~ n_i \in N$ and $\nu$ is the frequency of a normal mode of coupled oscillations. Depending on $\nu$, we can have $z < 1$. In such cases, ${\mathcal{Z}} = z^2 < z$, when two normal modes with the same frequency are available. However, such a probability distribution also indicates that 
    	${\mathcal{Z}_{N_l \times N_l}} = {z_l}^2 < {z_l}$, where $z_l$ is the corresponding partition function for the set of first $l - 1$ positive integers including zero and $l > 1$. Thus, in these cases the partition function itself does not give a faithful measure of $|N_l \times N_l|$ relative to that of $|N_l = \{ 0,1,2,..,l - 1 \}|$ for any finite value of $l$ although ${\mathcal{Z}_{N_l \times N_l}} = {z_l}^2$ indicates $|N_l \times N_l| = l^2$. It is more appropriate to use the internal energy and entropy as the measures of $|N_l \times {N_l}|$ relative to $|N_l|$ for such cases. In the following we will find the same continues to remain valid when $N_l$ is replaced by $N$. The requirement is the existence of a probability distribution on $N \times N$ which obeys: $p(|n_1 , n_2>) = p(|n_1>) p(|n_2>), ~n_i \in N$.

    	We now return to $g_{\vec{k}}(n_1,n_2, \beta)$ for which the single polarization partition function $z(\vec{k}, \beta)$ is greater than one, \textit{i.e}, $z > 1$. In this case, ${\mathcal{Z}_{N_l \times N_l}} = {z_l}^2 > {z_l}$ for any finite $l (> 1) \in N$. This agrees with: $|N_l \times N_l| = |N_l|^2 > |N_l|$ and ${\mathcal{Z}_{N_l \times N_l}}$ gives a faithful measure of $|N_l \times N_l|$, i.e, the size of the underlying sample space. To discuss the corresponding situation with $N$, we illustrate geometrically the second line of Eq.(4) by considering the lattice of points $(n_1,n_2), ~n_i \in N$, in a two dimensional plane. The expression in the third parentheses gives the sum of $\exp(-{\alpha}{n_{2}})$ over the column of points with coordinates $(n_1,n_2)$, where $n_1$ is fixed and $n_2 \in N$. $\mathcal{Z}(\vec{k}, \beta)$ is then obtained by multiplying the resulting expression with a weight factor $\exp(-{\alpha}{n_{1}})$ and summing over the row with coordinates $(n_1,0), n_1 \in N$. Thus, the second line of Eq.(4) and
    ${({{e^{\alpha}} \over {e^{\alpha} - 1}})^2} > {({{e^{\alpha}} \over {e^{\alpha} - 1}})}$ for $\alpha > 0$ imply that the cardinality of $N \times N$ is greater than that of $N$ itself: 
    
    \be
    |N \times N| = |N|^2 > |N|. 
    \ee
    
    \noindent{To justify further, we can replace $\exp(-{\alpha}{n_{1}})$ in the second line of Eq.(4) by $1$ to obtain
    	$\mathcal{Z'}(\vec{k}, \beta) = z(\vec{k}, \beta) |N|$. Although $|N|$ is given by a cardinal number, from $z(\vec{k}, \beta)l > z(\vec{k}, \beta)$ for any $l > 1, ~l \in N$ we conclude: $\mathcal{Z'}(\vec{k}, \beta) = z(\vec{k}, \beta) |N| > z(\vec{k}, \beta)$. The same remains valid for $0 < z <1$. It is easy to show this by considering $z(\vec{k}, \beta) l, ~l \in N$ and $l > 1/z(\vec{k}, \beta)$. These remarks imply $|N \times N| > |N|$. Replacing both $\exp(-{\alpha}{n_{1}})$ and $\exp(-{\alpha}{n_{2}})$ by $1$ in the second line of Eq.(4) gives Eq.(7). The later case corresponds to a microcanonical ensemble and will be discussed later. The above method of counting the microstates and hence evaluating the cardinality of $N \times N$ is confirmed by experiments. We will discuss this in the following. It is easy to generalize $g_{\vec{k}}(n_1,n_2, \beta)$ and $\mathcal{Z}(\vec{k}, \beta)$ to $N^m$ for finite $m \in Z_+$, and corresponding expressions imply that $|N^m| = |N|^m > |N|$. This is also consistent with $|{N_l}^n| = |N_l|^n > |N_l|$ for all finite $l$ and inductive arguments [5].}

    We now discuss the experimental verification of Eqs.(3,6,7) by considering two extensive thermodynamic variables, namely, the internal energy and entropy of the electromagnetic field with wave vector $\vec{k}$. These are additive thermodynamic variables that increase with the cardinality of the set of available microstates and are suitable to express corresponding cardinality. The internal energy is given by:

\ba
U(\vec{k}, \beta) & = &  {\sum_{n_1 \in N}} ~{\sum_{n_2 \in N}} {h\nu}(n_1 + n_2){p(|n_1,n_2>; \beta)}  
= 2 {\sum_{n \in N}}{n h\nu}p(|n>; \beta) = 2 h \nu <{n_{\vec k}}> = 2 u(\vec{k}, \beta) \\ \nonumber
& = &  - {{\partial} \over {\partial \beta}} \ln[\mathcal{Z}(\vec{k}, \beta)]  
= - 2{{\partial} \over {\partial \beta}} \ln[{z}(\vec{k}, \beta)] \\ \nonumber
<{n_{\vec k}}> & = & {{1} \over {e^{\beta h \nu} - 1}}
\ea

\noindent{where $<{n_{\vec k}}>$ gives the average occupation number of photons with any specific polarization and wave vector $\vec k$. $u(\vec{k}, \beta)$ is the corresponding internal energy. The above expression is experimentally verified by measuring the spectral radiance given by [12]:
	
	\[ R(\nu, \beta) = \frac{\nu^2 U(\vec{k}, \beta)}{c^2} = {\frac{\nu^2}{c^2}} {{2h \nu} \over {e^{\beta h \nu} - 1}} \]
So far, experiments have not contradicted the above expression of $R(\nu, \beta)$ and the underlying probability distribution given by Eqs.(3,4,6) [3,12]. Every term in the double sum of Eq.(8) is non-negative. Both polarizations contribute equally to $U(\vec{k}, \beta)$ and replacement of the double sum over $N \times N$ by twice a single sum over $N$ and, hence, the factor 2 in Eq.(8) demonstrates that $\{ |n_1,n_2>  |~ n_i \in N \}$ contains more microstates than $\{ |n> |~ n\in N\}$, \textit{i.e}, either $\{ |n,0> |~ n\in N\}$ or $\{ |0,n> |~ n\in N\}$. We conclude: $|N \times N| > |N|$. We can also arrive at the same conclusion only by using the experimental observation: $U(\vec{k}, \beta) = 2 u(\vec{k}, \beta)$. The second line shows that the corresponding normalization constants obey: $\mathcal{Z}(\vec{k}, \beta) = z^2 (\vec{k},\beta)$. $z(\vec{k},\beta)$ gives a measure of $N$.
With $z(\vec{k},\beta) > 1$, measurement of $R(\vec{k}, \beta)$ also experimentally confirms Eq.(7), i.e, $|N \times N| = |N|^2 > |N|$. The definition of internal energy given in terms of $p(|n_1,n_2>; \beta)$ in the first line of the above equation remains valid for an arbitrary probability distribution not necessarily described by a canonical ensemble. We continue to have: $U(\vec{k}) = 2 u(\vec{k})$, provided the joint probability distribution obeys: $p(|n_1 , n_2>) = p(|n_1>) p(|n_2>), ~n_i \in N$ and 
$p(|n _1>) = p(|n_2>)$, \textit{i.e}, single polarization probabilities are independent of the polarization. Thus, existence of such probability distributions on $\Omega$ indicates that $|N \times N| > |N|$. Eq.(6) gives an experimentally observed example. We obtain: $U(\vec{k}) = {u_1}(\vec{k}) + {u_2}(\vec{k})$, when $p(|n_1>) \neq p(|n_2>)$. This expression also indicates $|N \times N| > |N|$. The case of microcanonical ensemble with the set of microstates given by $\Omega$ requires special attention and will be considered later. We now consider the quantum entropy of the black body radiation.}

In general, the von Neumann quantum entropy or simply the entropy of a thermodynamic system is given by the following expression:

\be
S = - k_B \sum_{i} {p_i \ln(p_i)}
\ee 

\noindent{where the sum is over the accessible microstates to the system and $p_i$ is the probability of occupying the $i$-th state. In the present case, the microstates are given by $\Omega = \{ |n_1 , n_2> | ~n_1 , n_2 \in N \}$ with 
$p_i = {p(|n_1,n_2>; \beta)} = {g_{\vec{k}}(n_1 , n_2) / \mathcal{Z}(\vec{k}, \beta)}$, and we get:}

\ba
S(\vec k , \beta) & = & - 2 k_B {\sum_n} {{\exp(-{\alpha}{n})} \over \sqrt{\mathcal{Z}(\vec{k}, \beta)}}
\ln[{{\exp(-{\alpha}{n})} \over \sqrt{\mathcal{Z}(\vec{k}, \beta)}}], ~ n \in N \\ \nonumber 
~ & = & 2 s(\vec k , \beta)
\ea

\noindent{where $s(\vec k , \beta)$ is the entropy of the photon gas with any one of the two polarizations. We can also express the entropy by using the partition function and internal energy to derive:}

\ba
S(\vec k , \beta) & = & {k_B} \ln[\mathcal{Z}(\vec{k})] + {k_B} \beta U(\vec{k}) \\ \nonumber
~ & = & 2 {k_B} [\ln ({{e^{\beta h \nu}} \over {e^{\beta h \nu} - 1}}) + {{\beta h \nu} \over {e^{\beta h \nu} - 1}}] \\ \nonumber
~ & = & 2 s(\vec k , \beta).
\ea

\noindent{For a massless particle gas with $m$ polarization states, the microstates are given by: 
	$\{ |n_1 , n_2 , .. , n_m> | ~n_1 , n_2 , .. , n_m \in N \}$ and it is easy to find that: 
	
	\be
	S(\vec k , \beta)/s(\vec k , \beta) = m. 
	\ee
	
	\noindent{The above relation holds for statistically independent events, \textit{i.e}, whenever the joint probability distribution is product of individual probabilities: $p(|n_1 , n_2 , .. , n_m>) = p(|n_1>) p(|n_2>) .. p(|n_m>)$, where $n_i$ gives the number of photons with polarization $\vec{\epsilon}_i$. This was assumed in Eqs.(3,6).	For $m$- polarization states we have: $\ln[{\mathcal{Z}(\vec{k}, \beta)}] = m \ln{z(\vec{k}, \beta)}$ resulting in Eq.(12). Eq.(7) remains valid for any finite $N_l = \{0,1,2,..,l-1\}, l^2 > l, l > 1$.  We find that extending this result to $N$ itself, \textit{i.e}, taking the limit $(l - 1) \rightarrow \infty$ and replacing $l$ by $|N|$ agrees with the familiar interpretation of entropy as a measure of disorder in a thermodynamic system, \textit{i.e}, the cardinality of accessible microstates increases with increasing entropy. We can again extend Eq.(11) to the case of an arbitrary probability distribution where individual probability distribution is not necessarily given by a canonical ensemble but occupying $\vec{\epsilon}_1$ and $\vec{\epsilon}_2$ are statistically independent events
and corresponding single polarization probabilities are independent of the polarization. For the probability distribution given by Eq.(6), all of $\mathcal{Z}(\vec{k}, \beta)$, $U(\vec{k}, \beta)$ and $S(\vec k , \beta)$ are useful variables to express the cardinality of $N ^m$ relative to that of $N$. We also note that $||\ln[\mathcal{Z}(\vec{k}, \beta)]||$ and $||{{\partial} \over {\partial \beta}} \ln[\mathcal{Z}(\vec{k}, \beta)]||$ increase linearly with the order $'m'$ of the Cartesian product $N ^m$. This is consistent with the experimentally observed extensive character of the internal energy and entropy [3].}

We conclude the discussions on entropy by considering the microcanonical ensemble with $\Omega = {N_l}^m$. 
For a microcanonical ensemble, all states $|n_1 , n_2 , .. , n_m>$ are equiprobable and $p(|n_i>) = 1/l$. The single polarization states entropy is given by ${k_B}\ln(l)$.  We again have: ${U_l}/u_l = m$ and ${S_l}/s_l = m$, where $u_l$ and $s_l$ give the single polarization states internal energy and entropy respectively.
We next consider the microcanonical ensemble with $(l - 1) \rightarrow \infty$ and $N_l$ is replaced by $N$. $p(|n_i>)$ can be formally represented by $1/|N|$ which does not correspond to any real number. The entropy can be represented by the cardinal number [6]: ${k_B}{\ln(|N|)}$, and this also does not coincide with any real number. Similar comments remain valid for the internal energy. We will demonstrate in the next section that we can define $S = ms$ in this case also by extending the corresponding result valid for ${N_l}^m$. Cardinal numbers appear to describe a microcanonical ensemble when the set of accessible microstates is discrete and infinite. It would be more appropriate to use other definitions of entropy for such systems to express it using real numbers. Alternatively, we can choose $p(|n_i>) = 0, ~ i = 1,2,..,m$. However, both internal energy and entropy vanish for this choice which is inappropriate and $S = ms$ remains valid trivially. We conclude that when the cardinality of available microstates are given by cardinal numbers, formation of a statistical system from its microscopic constituents and thermalization [13] take place simultaneously.  This allows us to obtain non-zero values for thermodynamic variables like internal energy and entropy that can be expressed by real numbers. This is consistent with the quantum mechanical description of statistical systems that often uses infinite dimensional vector spaces.

	The total internal energy and entropy of the electromagnetic field in the enclosure at a temperature $T$ can be found by summing over all possible $\vec q$. To elaborate, the internal energy per unit volume in the limit of large volume is given by:
	
	\ba
	\mathcal{U} & = & \int_{0}^{\infty} d\nu \mathcal{U}(\nu, T) \\ \nonumber
	\mathcal{U}(\nu, T) & = & {{8 \pi h} \over {c^3}} {{\nu^3} \over {e^{\beta h \nu} - 1}}
	\ea

	\noindent{This expression agrees with the experiments [3,12] and confirms that Eq.(6) gives the correct probability distribution law of occupation on $\Omega$. We also conclude that both internal energy and entropy give valid measures of the sample space's size associated with every macrostate, \textit{i.e}, the cardinality of the underlying microstates given by $\Omega$ or $N \times N$. Thus, Eqs.(3-8,13) agree with the experiments and lead us to conclude that $N \times N$ gives greater amount of microstates than $N$ itself [3,11,12]. It is common practice to state that two transverse polarizations of the photons give two degrees of freedom. The cardinality of the $\Omega$ is expressed by $|\Omega| = |N|^f$, where $f = 2$ gives the degrees of freedom [3,13].}

		Eq.(7) is also consistent with Definition 1-1 and the stereographic projection discussed in the present section. The point $M$ with coordinate $(0,1)$ can't be assigned any  stereographic coordinate including the integers. Thus, according to Theorem 1-2, we expect the set $(0,1) \bigcup \{ (s_n,0)| ~n\in N \}$, to have a greater cardinality than that of $N$. Here, $s_n$ are the set of points on $S^1$ that are mapped to $(n,0), ~ n \in N$, on the $X$ -axis under the stereographic projection. This implies that the set $(0,1) \bigcup \{ (n,0)| ~n\in N \}$ can have a greater cardinality than $N$. We will prove a similar result in the next section.

	\section*{3. Cardinality of $Z_+ \times Z_+$ and $Q_+$}

	We commence our discussion on the cardinalities of $Z_+ \times Z_+$ and $Q_+$ with convergent sequences of rational numbers. The convergent sequence of positive rationals: $\{ r_{n} = 3/2 + 1/n \}, ~ n \in Z_{+}$ is isomorphic to $Z_{+}$ and has the same cardinality as that of $Z_{+}$ by the $Schr\ddot{o}der-Bernstein$ theorem. The set $G = \{ 1/2 \} \bigcup \{ r_{n} = 3/2 + 1/n \}, ~ n \in Z_{+}$, is larger than $\{ r_{n} =  3/2 + 1/n \}, ~ n \in Z_{+}$. We expect $|G| > |Z_+|$ when we assume both $G$ and $Z_+$ to be countable. It is possible to construct an injective function from $(-2,2)$ to $(-1,1)$ due to the order property $O_2$ of $R$ and both sets are uncountable. However, $Z_{+}$ does not possess the order property $O_2$ of $R$ and the only way to construct an injective function from $G$ to $Z_{+}$ is by proposing integers of the form $\aleph = n + 1, ~n \rightarrow \infty$. Such an integer can't exist since it implies the existence of a real number: $z = x + \epsilon, ~ \epsilon > 0, ~ x \rightarrow \infty$. The later is forbidden by Definition 1-1. Alternatively, the limit $n \rightarrow \infty$ implies $n$ is increased without any upper bound and existence of the integer $\aleph$ leads to problems similar to those mentioned above Definition 1-1. This is also consistent with the least limit ordinal character of $N$ containing the natural numbers, that comes next to all natural numbers [6]. To illustrate, we consider the sequence of sets: 
	
	\[ 0 = \emptyset, ~~1 = \{0\}, ~~2 = \{0,1\}, ~~3 = \{0,1,2\}, ~~...  ~.\] 
	
	\noindent{Thus, after all the natural numbers comes the set: $N = \{0,1,2,3,...\}$, [6]. Taking the limit $n \rightarrow \infty$ means we have used up all the natural numbers and the existence of $\aleph$ as a natural number violates the existence of $N$ as the least limit ordinal containing all natural numbers. This objection persists whenever we construct any injective function $f(n)$ from $Z_{+}$ to $Z_+$ such that $f(n)$ is finite for finite values of $n$ but $f(n) > n$ as $n \rightarrow \infty$. This happens for: $2n - 1, 2n, {m^n}; ~ m,n \in Z_{+}$ and $m$ is finite. These functions exist only for any finite value of $n$. In particular, the function ${2^m}{3^n}$ is not well-defined on $Z_+ \times Z_+$ in the limit when one or both of $m,n \rightarrow \infty$. Such functions are required to construct injective functions from $Z_+ \times Z_+$ to $Z_+$,[5,6]. 
	Thus, we conclude that $Z_+ \times Z_+$ need not to be countable. It is easy to construct an injective function from $Z_l \times Z_l$ to $Z_+$ but there cannot exist any injective function from $Z_l \times Z_l$ to $Z_l$, where $l$ is a finite positive integer and $Z_l = \{1,2,..,l\}$ is the set of first $l$ positive integers. For $N$, $|N_l \times N_l| > |N_l|$ for any finite $l > 1$ where $N_l = \{ 0,1,2,..,l - 1 \}$. We have found in the previous section that the same remains valid in the limit $(l - 1) \rightarrow \infty$, so that $N_l = N$, in the counting procedure adapted to count the microstates of black body radiation. This is also consistent with the inductive arguments.}

	We now consider the set of positive rationals $Q_+$. When we count the number of 
	elements in a countable set $A$, we first separate the elements of $A$ and enumerate them 
	in an increasing order using the elements of $Z_{+}$ as $1,2,3..$ without any gap. 
	The last element in the above sequence gives the cardinality of $A$. We can use the Hausdorff topology of the real numbers [5] to separate and enumerate the elements of a set similar to $G$ mentioned above. We now prove the following theorem. A preliminary version of the proof can be found in [2]. We follow an approach similar to the proof of Theorem \textbf{3}-6.5., [5]. We first note that a bijective function on $Z_{+}$ is used to construct a sequence $\{ x_{n} \} ~(n \in Z_{+})$ isomorphic to $Z_{+}$. A different bijective function from $Z_{+}$ to the same sequence $\{ x_{n} \}$ can be regarded as a permutation on the elements of $Z_{+}$ because the index set of $\{ x_{n} \}$ is $Z_{+}$.

\vspace{0.5cm}

\textbf{Theorem 2-1.} \textit{The set of positive rational numbers and the set of all rational numbers are uncountable.}

\vspace{0.2cm}

\textit{Proof}. We will first construct a set $A = S \bigcup X$, where $S$ is a sequence of
positive rationals isomorphic to $Z_{+}$ and $X$ is another set of positive rationals whose
elements are different from $S$. We will consider $X$ to be finite. 
Thereafter, we will show that $|Z_{+}| < |A|$. It then follows that: $|Z_{+}| < |Q_{+}|$.

We consider a nested family of closed intervals of $R$ given by: $[3/10, {1/2}], [33/100, {1/2}], [333/1000, {1/2}],
...$. We index these intervals as $I_{1}, I_{2}, I_{3},...$, with $I_{n}$ containing $I_{n + 1}$
and $n \in Z_{+}$. We now consider a set of points ${y_{n}} \in R$. For $n \geq 2$, $y_{n}$ is the midpoint between 
the lower boundaries of $I_{n - 1}$ and $I_{n}$ and is given by 
$y_{n} = ({1 \over 2})[{{x_{1}....x_{n - 1}} \over {10^{n - 1}}} + {{x_{1}....x_{n}} \over {10^{n}}}]
= (1/2)(0.x_{1}....x_{n - 1} + 0.x_{1}....x_{n})$,
where all $x_{i}$ are $3$ and ${y_{1}} = 3/20 = 0.15$. All ${y_{n}}$ are different and have rational coordinates. 
The first few are given by ${3/20}, {63/200}, {663/2000}, {6663/20000}$. By the Housdorff property of the standard topology of $R$, 
the lower boundary of every $I_{n}$ can be separated from 
the corresponding (in index) $y_{n}$ by two disjoint open intervals and every $I_{n}$ 
excludes all ${y_{m}}$ with $m \leq n$. We can construct the bijective function $f: Z_{+} \rightarrow \{ y_{1}, y_{2}, ... \}$ 
by $f(n) = {y_{n}}, ~ n \in Z_{+}$. Each $I_{n}$ excludes the corresponding point ${y_{n}}$. 
We now consider the closed interval $I_{\aleph} = [{1/2.9}, {1/2}]$. $1/3$ is the limit of
the sequence $\{ u_{n} = {x_{1}....x_{n} \over {10^n}} \}, ~ n \in Z_{+}$ and ${x_{i}} = 3$ for all $i \in Z_{+}$. 
$I_{\aleph}$ is contained in the intersection of all $I_{n}$, contains the points 
with coordinates ${1/2.9}$ and ${1/2}$, but none of $f(n) = {y_{n}}$ is contained in $I_{\aleph}$. 
We find that $I_{n}$ can be arbitrarily close to $[{1/3}, {1/2}]$ but can never be $I_{\aleph}$.
We can replace $\aleph$ by the cardinal that represents the cardinality of $Z_{+}$.
We continue to use $\aleph$ with the note that $\aleph \notin Z_{+}$.

We now consider the set $A = \{ y_{1}, y_{2}, ... \} \bigcup \{ {y_{\aleph}} = 1/2 \}$.
The function $f$, although injective, is not a surjection of $Z_{+}$ to $A$. 
There cannot exist such a surjection from $Z_{+}$ to $A$ since $\aleph \notin Z_{+}$.
If exists, any one-to-one surjective (bijective) function $g$ from $Z_{+}$ to $A$ can be considered to be a permutation on the elements of $Z_{+}$
onto the index set of the elements of $A$. This can be easily understood if we compose with $g$ from the left 
the bijective function $h$ from $A$ to the index set of its elements where, 
$h(y_{n}) = {f^{-1}(y_{n})} = n, ~ n \in Z_{+}$ and $h({y_{\aleph}} = 1/2) = \aleph$.  
Such a permutation is possible if the index set of $A$ would have been $Z_{+}$. 
However, we cannot permute the elements of $Z_{+}$
to generate an additional element so that they can correspond to the elements of a set
containing one more element in addition to those present in $Z_{+}$. 
Thus, there cannot exist a one-to-one surjective function from $Z_{+}$ to $A$.
As mentioned before, it is important to keep in mind that there exists no injective 
function $f(n)$ from $Z_{+}$ to $Z_+$ such that $f(n)$ is finite for finite values 
of $n$ but $f(n) > n$ as $n \rightarrow \infty$. This is relevant to construct $h$ given above and also to permute the elements of $Z_+$. We can consider other examples of $S$ and $I_{n}$. We can take $I_{1} = [1/4, 3/2], ~ z_{1} = 8/5 = 1.6$ and $I_{n} = [1/4 ,{1/2 + {1/n}}], ~ z_{n} = {1 \over 2}  + ({1 \over 2})[{1 \over {n - 1}} + {1 \over {n}}]$
for $n \geq 2$ and $n \in Z_{+}$. We take $I_{\aleph} = [1/4,1/2]$. 
In this case, the point with coordinate $1/2$ is the limit of the sequence $\{ z_{n} \}, ~ n \in Z_{+}$.
We replace $A$ by $B = \{ z_{1}, z_{2}, ... \} \bigcup \{ z_{\aleph} = 1/4 \}$.
Similar arguments as above show that there cannot exist a one-to-one surjective (bijective) function from $Z_{+}$ to $B$.

We now show that there cannot exist a many to one  surjective function from $Z_{+}$ to $A$. 
To elaborate, let there is a surjective function $p$ from $Z_{+}$
to $A$ which is not injective. Let $Y$ be the subset of $Z_{+}$ that is not mapped injectively
to $A$. Let, $p(Y) = P; ~Y \subset Z_{+}, ~P \subset A$ and $|P| < |Y|$. $P$ is a proper subset of $A$ and
$Y$ cannot be $Z_{+}$ if $p$ has to be surjective. For any such $p$, we can 
always redefine the bijective function $f$ used before in the proof to
another bijective function $f': Z_{+} \rightarrow \{ y_{1}, y_{2}, ... \}$ 
so that $f'(Z_{+} - Y)\bigcap P = \emptyset$. We cannot restrict $p$ to 
a bijective function from $Z_{+} - Y$ to ${[f'(Z_{+} - Y)]} \bigcup {[f'(Y) \bigcup \{ y_{\aleph} \} - P]}$.
The arguments are similar to those used above to prove the corresponding result for $Z_{+}$ and $A$. We should note that 
${f'(Y) \bigcup \{ y_{\aleph} \} - P}$ is nonempty and is different from ${f'(Z_{+} - Y)}$.  
They have different index sets.

We next show that we cannot have $Y = Z_{+}$ and $P = A$.
In this case, we can restrict $p$ to construct a bijective function $p'$ from a suitable subset of $Z_{+}$
to $A$. $(p')^{-1}$ gives us an injective function from $A$ to $Z_{+}$. 
We exclude these cases of $p$ by showing that we cannot have an injective function from $A$ to $Z_{+}$. 
If exists, every injective function $q$ from $A$ to $Z_{+}$
can be considered to be a bijection $q'$ on the indices of $\{ y_{n} \}, ~ n \in Z_{+}$ and ${y_{\aleph}}$
to the image set $q(A)$. We can use the bijective function ${h^{-1}}$ on the index set of $A$ and compose it
with $q$ from the right $(= q{h^{-1}})$ to understand this. 
A bijection on the elements of a set cannot give a lesser number of images.
Since $\aleph \notin Z_{+}$, it is not possible to have an injective function from $A$ to $Z_{+}$
even if the image set of $q$ coincides with $Z_{+}$. 
Similar conclusion is valid for the second example with $A$ replaced by $B$.
Thus, there cannot exist a surjective function from $Z_{+}$ to $A$ or $B$.

It follows from Theorem 1-2 that $|Z_{+}| < |A|$ and $|Z_{+}| < |B|$.
We conclude that $A$ and $B$ are uncountable.
$A$ and $B$ are proper subsets of $Q_{+}$. Thus, $Q_{+}$ has a higher cardinality than $Z_{+}$ and
we conclude that $Q_{+}$ is uncountable. It is obvious that the set of all rational numbers is also uncountable. Q.E.D.

\vspace{0.5cm}

From the set theoretic perspective, the non-integral rationals are as different from the integers as the 
irrationals are compared to the rationals. This is in particular significant for counting. 
It is more appropriate to have two different cardinal numbers [6] to represent the 
cardinal properties of $Z_{+}$ and $Q_{+}$ if $Z_+$ is considered to be the largest countable set. We have shown in the above theorem that $Q_{+}$ has a greater
cardinality than $Z_{+}$ using Definition 1-1. Existence of the irrationals indicates that we have the hierarchy
$|Z_{+}| < |Q_{+}| < |R_{+}|$. According to Cantor's theorem on the cardinality of power sets,
the power set of $Z_{+}$ given by $P(Z_{+})$ has a greater cardinality than that of $Z_+$ [6].
The positive rationals can be represented by a proper subset of $P(Z_{+})$ that consists of subsets of $P(Z_{+})$ having single element, subsets of $P(Z_{+})$
having two elements and subsets of $P(Z_{+})$ having three elements. The last subsets are required
to distinguish between rationals like $2/7$ and $7/2$. Thus, we conclude that $|Z_{+}| < |Q_{+}| < |P(Z_{+})|$ 
if we assume $|R_{+}| = |P(Z_{+})|$, [5]. This gives us a departure from the \textit{continuum hypothesis} [5]. $Z_{+}$ is the least limit ordinal of the set of positive integers [6]
and we now introduce the cardinal numbers ${\varepsilon} = |Z_{+}|, {\varphi} = |Q_{+}|$ and ${\varpi} = |R_{+}|$.
Theorem 2-1 indicates that ${\varepsilon} + n > {\varepsilon}, ~ n \in Z_{+}$. 
Theorem 2-1 also indicates that the Cartesian products $Z_{+}^{n}$ are not countable
and ${\varepsilon}^{n} > {\varepsilon}$. This is consistent with the experimental observations
from statistical physics mentioned before. Thus, ${\varepsilon}$ is similar to the positive integers in these
aspects. We can also state that $2^{\varepsilon} > {\varepsilon}$.
This is good for ${\varepsilon}$ to represent the cardinality of a set since the cardinality of the power set of a set 
is always greater than that of the set.  These conclusions give us an improvement over the conventional literature [5]. However, ${\varepsilon}$ cannot be obtained from the integers inductively
and we can't obtain ${\varphi}$ or ${\varpi}$ from ${\varepsilon}$ inductively.
We also use the term cardinal numbers to denote cardinalities like ${\varepsilon} + n$ and ${\varepsilon}^{n}, ~ n \in Z_{+}$, discussed above. We will later include cardinalities like ${\varepsilon} - n, ~ n \in Z_{+}$
in the set of cardinal numbers. We can add $0$ to the respective sets to conclude that $|N| < |Q_0| < |R_0|$,
where $Q_0 = \{ 0 \}\bigcup Q_{+}$ and $R_0 = \{ 0 \}\bigcup R_{+}$. 
It is easy to show that $|Z_{-}| < |Q_{-}| < |R_{-}|$ by using Theorems 2-1 and 1-1. Thus, we find that $|Z| < |Q| < |R|$, where $Q$ is the set of all rational numbers. 
This is expected since $Q$ and $R$ satisfy the order property $O_{2}$ but $Z$ does not. Lastly, for $R^n$ where $n \in Z_+$ and finite, it is more appropriate to introduce a metric and use volume to compare the sizes of sets homeomorphic to $R^n$.

We now resume our discussions on the entropy of black body radiation by considering the case where $\beta = 0$ in Eq.(3). Despite being physically unattainable, we obtain a microcanonical ensemble in which all accessible microstates are equiprobable. We first replace $N$ by $N_l = \{0,1,2,..,l-1\}$. The entropy is now given by: ${S_l}(\vec k) = k_B \ln(|N_l \times N_l|) = k_B \ln(l^2) = 2 k_B \ln(l) = 2 k_B \ln(|N_l|) = 2 {s_l}(\vec k)$, and we again find that ${S_l}(\vec k) / {s_l}(\vec k) = 2 = \ln(|N_l \times N_l|)/\ln(|N_l|)$. Extending the last expression to the limit $(l - 1) \rightarrow \infty$ gives:} 

\ba
S(\vec k)/s(\vec k) & = & 2 \\ \nonumber
\implies \ln(|N \times N|)/\ln(|N|) & = & 2. 
\ea

\noindent{Here, $|N|, \ln(|N|), S(\vec k)$ and $s(\vec k)$ are given by cardinal numbers. $|N|$ can be given by ${\varepsilon} + 1$ and $\ln(|N|)$ refers to the cardinal number $\vartheta$ such that ${\varepsilon} + 1 = \exp(\vartheta)$, the later being similar to $2^{\varepsilon}$. The above expressions formally agree with Eq.(10) and Eq.(7) respectively. The later were derived for the experimentally validated canonical ensemble with $S(\vec k , \beta)$ and $s(\vec k , \beta)$ given by ordinary real numbers. The ratio $S(\vec k)/s(\vec k)$ in Eq.(14) is given by an ordinary integer and aligns with the interpretation of entropy as giving an estimate of the amount of microstates available to a given thermodynamic system. Thus, we have defined the ratio between the logarithm of two cardinal numbers $|N \times N|$ and $|N|$ by extending the corresponding result for ordinary real integers. We conclude that
entropy and internal energy of a statistical system can provide expressions consistent with the Definition 1-1 and Theorem 2-1 to measure the cardinality of the underlying set of microstates.}

It is always better to use bijective functions
to resolve confusions regarding the limit ordinals. To illustrate, we can obtain the set $Z_{+} - \{ 1 \}$ inductively from $Z_{k} - \{ 1 \} = \{ 2,3,..,k \}$ for any finite $k \geq 2$. It is now easy to show that
$(Z_{+} - \{ 1 \}) \notin [Z_{+}]$. $Z_{+} - \{ 1 \}$ can be considered 
to be the least limit ordinal of the set of all positive integers excluding $1$. 
$Z_{+} - \{ 1 \}$ is countable. Any element of $Z_{+} - \{ 1 \}$ gives
the cardinality of a countable set except when the set contains a single element or the set is isomorphic to $Z_{+}$.
We represent the cardinality of $Z_{+} - \{ 1 \}$ by the cardinal number $\varepsilon - 1$. 
$\varepsilon - 1$ is less than $\varepsilon$ although we cannot represent it by any positive integer.
Similarly, we can represent the cardinality of the set of all odd integers or that of the set of all even integers 
by ${\varepsilon / 2}$. We can consider algebraic operations on cardinal numbers and non-negative integers as set-theoretic operations on appropriate members of the corresponding equivalence classes, whose cardinalities are represented by cardinal numbers and non-negative integers, respectively. Thus, we denote the cardinality of the set of all integers including zero by $2 \varepsilon + 1$, [2].

As a significant consequence of this article, Theorem 2-1 shows that the collection of open balls with rational radii centred at points with rational coordinates no longer provides a countable basis for the standard topology in $R$ [5] and is not a countable covering of $R$. The Lindel$\ddot{o}$f covering theorem in $R$ uses the above covering to conclude that $R$ is a Lindel$\ddot{o}$f space [14]. Thus, we need to construct a new countable covering to prove that $R$ is a Lindel$\ddot{o}$f space [14].

We can use the \textit{axiom of choice} to prove theorems that use countable basis [7,15]. A related issue is the metrizability and existence of connections in the spacetime manifold [7,8]. This corresponds to the transition from pregeometry to geometry [16]. Countability of the rational numbers gives rise to the concept of completely separable topological spaces (second-countable spaces) [7], which possess a topology with a countable basis, convenient for the introduction of metrics (Theorem 2-67 [7], Appendix 2 [8]). A locally compact and completely separable Hausdorff space is paracompact (Theorem 2-66 [7]). The later property allows to introduce connections globally in a topological manifold (Theorem 2.11 Chapter 2 [8]). On the other hand, Eqs.(7,8,10,13) follow from quantizing classical electrodynamics and various experimental observations establish general relativity based on connection and metric in the spacetime manifold which is locally $R^n$ [16,17,18]. This leads us to the \textit{axiom of choice} as an alternate tool to prove theorems that otherwise use the existence of a countable basis in a topological space (second-countable spaces). The present article indicates that the \textit{axiom of choice} can play a significant role in the differential topology of spacetime manifold and in the formal theories of quantum gravity, both of which require connection and metric [19,20]. More generally, present day quantum theory of particle physics is described by gauge theories that includes gravity [21,22]. Interactions are described by various connections where gauge invariance and stability might play significant roles to determine the actions [19,21] and particle spectrum [21]. The mathematics of transition from pregeometry to a geometry based on connections in the spacetime manifold can play a pivotal role to predict and explain the structure of relativistic and non-relativistic quantum theory of gauge fields. Quantum theory of gauge fields incorporates both propagators and entangled quantum states.

Lastly, we conclude the present article with the note that a covering using $Q^n, ~n \in Z_+$, is not troublesome when we don't require countability of $Q$ contradicting Theorem 2-1. In such cases, we can replace the term \textit{countable covering} by $R^n$-\textit{covering} for the collection of open balls with rational radii centred at points with rational coordinates in $R^n$.

\section*{Conclusion}

To conclude, we have demonstrated that the cardinality of the Cartesian products $N^m, m \in Z_+$ is greater than that of $N$ itself when $m \geq 2$. The same remains valid for $Z_+$. We have illustrated this by using the internal energy and  von Neumann quantum entropy of black body radiation. We have also found that when the cardinality of available microstates are given by cardinal numbers, formation of a statistical system from its microscopic constituents and thermalization take place simultaneously. This allows us to use ordinary real numbers to express non-zero values of thermodynamic variables like internal energy and entropy. We have demonstrated that the set of rational numbers is not countable and does not give a countable covering of $R^m, m \in Z_+$ for finite $m$. Thus, we find that the axiom of choice is a more rigorous technique to prove the existence theorems for metric and connection in the spacetime manifold. This also remains valid for the metrization theorems in a general topological space.

\vspace{2cm}

Conflict of interest statement: There is no known conflict of interest.

\vspace{0.5cm}

Data availability statement: My manuscript has no associated data.

\section*{References}

[1] K. Ghosh, \textit{Entropy and Cardinality of the Rational Numbers}, Talk given at the 2023 \textit{International Conference on Topology and its Applications}, July 3-7, 2023, Nafpaktos, Greece; https://hal.science/hal-05242734.

[2] K. Ghosh, J. Phys.: Conf. Ser. 2090, 012037	(2021), https://hal.archives-ouvertes.fr/hal-03092015.

[3] K. Huang, \textit{Statistical Mechanics}, Wiley-India Ltd., New Delhi, 2003.

[4] N. Jacobson, \textit{Basic Algebra I}, Dover Publications, Inc., New York, 1985.

[5] James R. Munkres, \textit{Topology A First Course}, Prentice-Hall of India Private Limited, 1994.

[6] Paul J. Cohen, \textit{Set Theory and the Continuum Hypothesis}, Dover Publications, Inc., 1994.

[7] John G. Hocking and Gail S. Young; Topology (Dover Publications, Inc., New York, 1961).

[8] S. Kobayashi and K. Nomizu, \textit{Foundations of Differential Geometry:} Vol. \textbf{I}, 
Wiley Classics Library, 1991.

[9] R. Courant and F. John, \textit{Introduction to Calculus and Analysis:} Vol.\textbf{1}, Springer-Verlag New York Inc., 1989.

[10] K. Ghosh, International Journal of Pure and Applied Mathematics,
\textbf{76}, No.2, pp.251 - 260, (2012).

[11] C. Itzykson and J. B. Zuber, \textit{Quantum Field Theory}, Dover Publications, Inc. Mineola, 2005.

[12] G. B. Rybicki and A. P. Lightman, \textit{Radiative Processes in Astrophysics}, John Wiley $\&$ Sons, 1979.

[13] Richard C. Tolman, \textit{The principals of Statistical Mechanics}, Dover Publications, Inc., New York, 1979.

[14] Tom M. Apostol, \textit{Mathematical Analysis}, Narosa Publishing House, 1992.

[15] E. Mendelson, Arch. math. Logik Grundlagenforsch 4 (1948), pp. 65-70.

[16] C. W. Misner, K. S. Thorne and J. A. Wheeler, \textit{Gravitation}, W.H. Freeman and company, New York, 1970.

[17] S. W. Hawking and G. F. R. Ellis, \textit{The Large Scale Structure of Space-Time}, Cambridge University Press, 1973.

[18] R. M. Wald, \textit{General Relativity}, The University of Chicago Press, Chicago and London, 1984.

[19] D. Lovelock and H. Rund, \textit{Tensors, Differential Forms, and Variational Principals}, Dover Publications, Inc., New York, 1989.

[20] K. Ghosh, Quantum Stud.: Math. Found. \textbf{11}, 625–642 (2024). https://doi.org/10.1007/s40509-024-00340-9.

[21] C. Quigg, \textit{Gauge Theories of the Strong, Weak and Electromagnetic Interactions}, Princeton University Press, 
Princeton and Oxford, 2013.

[22] Y. Aharonov and D. Bohm, \textit{Significance of Electromagnetic Potentials in the Quantum Theory}, The Physical Review. 115, No. 3, 1959, pp. 485-491.

\end{document}